\documentclass[referee]{raa}            
\usepackage{graphicx,times}             
\usepackage{natbib}
\usepackage{amssymb,amsmath}
\usepackage{color}
\usepackage{ulem}
\usepackage[dvipsnames]{xcolor}
\RequirePackage{lineno}
\usepackage{cancel}
\usepackage[OT1]{fontenc}

\bibpunct{(}{)}{;}{a}{}{,}
\usepackage[pagebackref=true]{hyperref}
\usepackage{siunitx}

\begin{document}

  \title{ An Exposure Meter of Lijiang Fiber-fed High Resolution Spectrograph 
}

   \volnopage{Vol.0 (20xx) No.0, 000--000}      
   \setcounter{page}{1}          

   \author{Xiao-Guang Yu 
      \inst{1,2,3}
   \and Kai-Fan Ji 
      \inst{1,2}
   \and Xi-Liang Zhang
      \inst{1,2}   
   \and Liang Chang 
      \inst{1,2,3}
   \and Yun-Fang Cai 
      \inst{1,2}
   \and Ying Qin
      \inst{1,2,3}
    \and Zhen-Hong Shang
      \inst{4}
    }

   \institute{Yunan Observatories, Chinese Academy of Sciences,
             Kunming 650011, China; {\it changliang@ynao.ac.cn; yuxiaoguang@ynao.ac.cn}\\
        \and 
            Key Laboratory for the Structure and Evolution of Celestial Objects, Chinese Academy of Sciences, Kunming 650216, China
        \and
            University of Chinese Academy of Sciences, Beijing 100049, China \\           
        \and 
        Faculty of Information Engineering and Automation, Kunming University of Science and Technology, Kunming 650500, China\\
   {\small Received 20xx month day; accepted 20xx month day}}

  \abstract{
    In 2016, an exposure meter was installed on the Lijiang Fiber-fed High-Resolution  Spectrograph to monitor the coupling of starlight to the science fiber during observations. Based on it, we investigated a method to estimate the exposure flux of the CCD in real time by using the counts of the photomultiplier tubes (PMT) of the exposure meter, and developed a software to optimize the control of the exposure time. First, by using flat-field lamp observations, we determined that there is a linear and proportional relationship between the total counts of the PMT and the exposure flux of the CCD. Second, using historical observations of different spectral types, the corresponding relational conversion factors were determined and obtained separately. Third, the method was validated using actual observation data, which showed that all values of the coefficient of determination were greater than 0.92. Finally, software was developed to display the counts of the PMT and the estimated exposure flux of the CCD in real-time during the observation, providing a visual reference for optimizing the exposure time control.}
   \authorrunning{X.-G. Yu et al}            
   \titlerunning{Exposure Meter of Lijiang Fiber-fed High Resolution Spectrograph}           
   \maketitle

   \section{Introduction}           
\label{sect:intro}

The signal-to-noise ratio (S/N) is one of the key parameters to evaluate the data quality of high-resolution spectra. During observations, because we can't obtain the CCD exposure flux information directly, the exposure time is generally estimated by considering the weather and observing conditions, and the characteristics of stars, etc, which has large uncertainty. If the exposure time is blindly increased for high S/N, it is a waste of observation time, and it may pollute the observed data because the CCD is saturated. If the S/N doesn’t meet the requirements, repeated observations will affect the observation efficiency.
There are usually two methods to solve the problem. One is using the Exposure Time Calculators (ETC), which needs to accurately simulate the target, weather and instrument information before observations. Most instruments use ETC to estimate exposure time, although it has some uncertainties. The simulated S/N from the ETC is overestimated by 40-50$\% $ on IGRINS \citep{etcIGRINS}, while it is within 10$\%$ for HARPS \citep{ESO}.

Another way is using Exposure Meter (EM), which is installed as a part of the instrument. the EM can monitor the flux in real-time to optimize the exposure time, because it can terminate the exposure when the desired counts have been acquired. The EM can also calculate the photon-weighted mean time of each exposure, which is critical for precise RV measurements \citep{FIES,CHIRON}. For precise RV measurements with long exposures, EM even becomes a necessity \citep{PHDTHermes}. Many high-resolution spectrographs have been equipped with EM, and the detectors of EM are photomultiplier tubes (PMT) or CCD \citep{KiwiSpec,HERMES,CHIRON,ESPRESSO,G-CLEF,EXPRES,HIRES&APF,NEID}. EM usually picks up from 1$\%$ to 8$\%$ of the light from behind the slit or grating. According to the different spectrographs, the count rates of EM are different. For a star of V=5mag., the typical count rate is \SI{103}{s^{-1}} in the EM of CHIRON \citep{CHIRON}, although its maximum count rate is \SI{456}{s^{-1}}, and the error of the mean exposure time is within 1s. However, for a star of V=16mag., the count rate is \SI{19}{s^{-1}}, and the error of the mean exposure time is within 165s for the EM of HARPS \citep{HARPS-manual}.

In this paper, we propose a method to estimate the exposure flux of a CCD in real-time using the counts of PMT in Fiber-fed High-Resolution Spectrograph (HiRES) of the Lijiang 2.4m telescope. Through flat-field lamp observation, we determined that the signals between the two detectors are linearly proportional, and obtained the relationship conversion coefficients for six different spectral types using historical observed data. The validation results from the observed data show that the estimated values of the CCD exposure flux are in good agreement with the actual observed values. At the same time, the application software was developed to realize the real-time display of the estimated CCD exposure during observation. Section 2 describes the principle of the method. Section 3 shows the validation results, Section 4 describes the working principle of the software and shows the user interface, and Section 5  summarizes the results of this work and discusses the error factors.


\section{LINEAR MODELING for Exposure Meter}
\label{sect:Obs}
\subsection{Optics of HiRES and Exposure Meter}

\begin{figure*}
  \centering
  \includegraphics[width=0.70\textwidth, angle=0]{{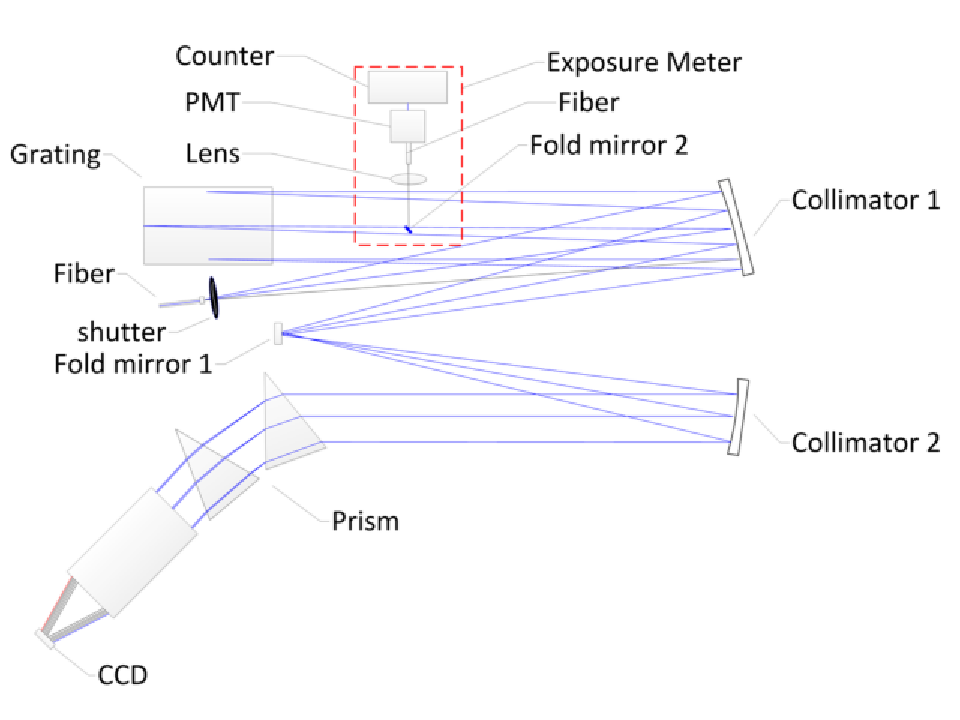}}
  \caption{ The optical system of HiRES for 2.4m Telescope,} the red part is photonic detection system. 
  \label{optics}
  \end{figure*}

HiRES of the Lijiang 2.4m telescope has two science fibers, the spectral resolutions($R=\lambda/\Delta\lambda$ ) are 32000 and 49000 for the 2.0” and 1.2” diameter fibers (on the sky) at 550nm, respectively. The orders of an observed spectrum are from 61 to 156, and the wavelength coverage is from 380 nm to 990 nm.  The long-term stability of temperature and pressure for the spectrograph are controlled within 26±0.25°C and 30±1Pa, respectively \citep{2m4Wcj}. A starlight coupling device was installed on a side port in the AG-box unit \citep{FYF} at the Cassegrain focus of the telescope and guided the light from the telescope or calibration lamps to the spectrometer via scientific fiber. Fig. \ref*{optics} shows the optical layout of HiRES of 2.4m telescope.

\begin{figure*}
  \centering
  \includegraphics[width=0.80\textwidth, angle=0]{{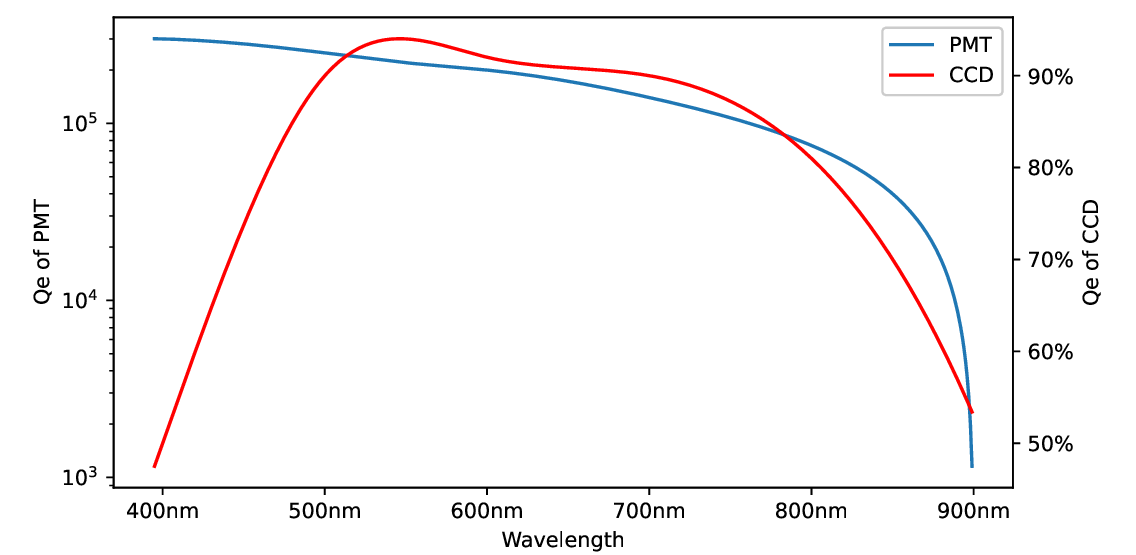}}
  \caption{Distribution of detection efficiency of PMT (model: Hamamatsu H8259-02) and CCD (model: E2V CCD203-82). }
  \label{qe}
  \end{figure*}

An EM (red box in Fig. \ref*{optics}) was installed in the spectrograph optical path for monitoring starlight coupling into science fiber during observation to pick up 3$\%$ starlight by a small folding mirror to EM, and its main component is a wide dynamic range PMT. 
Fig. \ref*{qe} shows the quantum efficiency curves of the two detectors (CCD and PMT), they have high detection efficiency in the wavelength range from 500nm to 800nm, and the peak quantum efficiency of PMT and CCD detectors are 400nm and 550nm, respectively. We used the observed data to study the relationship between the counts on EM and the flux on CCD of spectrograph, and developed a software to improve the observation efficiency. 

\subsection{Mathematical model}

\begin{figure*}
  \centering
  \includegraphics[width=0.8\textwidth, angle=0]{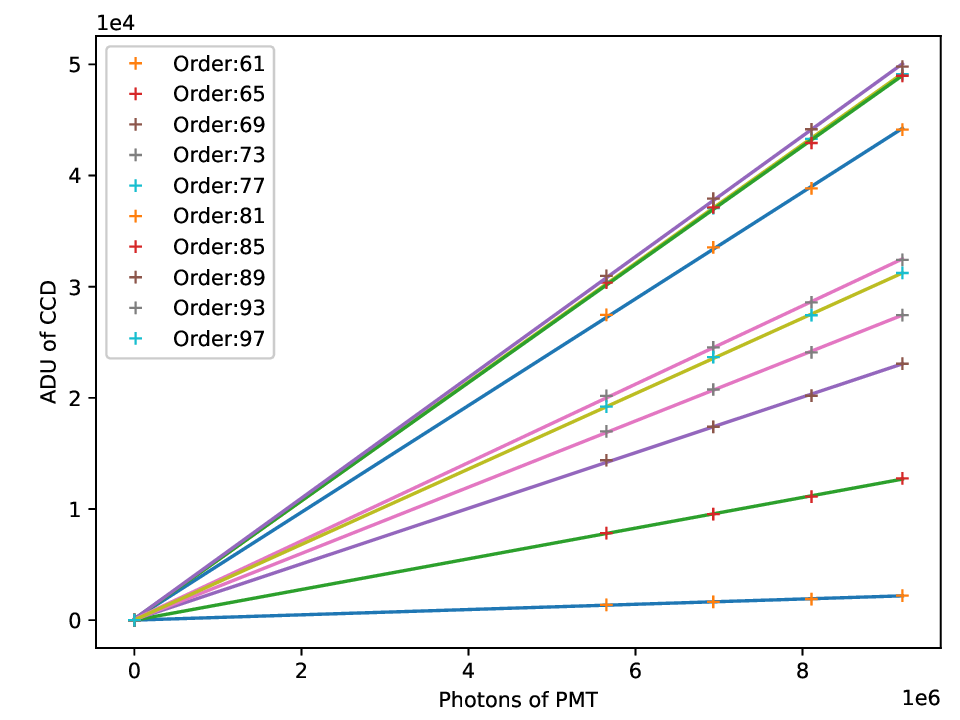}
  \caption{The linear fitting between the observed flux on CCD and the counts of PMT, the scatters represent the sampled values of the observed flux of different exposure time.}
  \label{fpmt2ccd}
  \end{figure*}

We first investigated the relationship between the signals from the two detectors (CCD and PMT). This was done by using flat-field lamp observations. Specifically, different exposure times (0.5s,0.6s,0.7s,0.8s due to the bright enough flat-field lamp) were set for group observations, and four sets of spectral images of the CCD and the corresponding total counts of the PMT were obtained.
First, the background signals of the spectral images and PMT counts are corrected.
Second, the 2000th column of pixels with the strongest signal in the spectral image was taken, and the ADU values of the pixels covered by the 96 orders in this column were summed individually to obtain the exposure fluxes at the wavelength points corresponding to the different orders. Fig. \ref*{fpmt2ccd} shows the relationship between the exposure fluxes of the 10 orders and the corresponding total PMT counts. The horizontal coordinates represent the total PMT counts and the vertical coordinates represent the exposure fluxes of the different wavelength points in the column of the CCD image. 

The dark counts for the CCD and the PMT were obtained before observations. In the no-exposure state, the effective signals of both detectors are zero. The relationship can be expressed in the form of the equation \ref*{eq1}. 
Where $F_{CCD}$ is the observed flux at the corresponding wavelength point for each pixel of the CCD, $C_{PMT}$ is the sum of counts of the PMT during the exposure time T, $G$ is the scale factor matrix of the corresponding pixel on CCD image, $d_{T}$ is the dark of the CCD during the exposure time $T$, $b_{T}$ is the sum of the dark counts of the PMT during the exposure time $T$.

\begin{equation}\label{eq1}
  F_{CCD} - d_{T} = G (C_{PMT} - b_{T})
\end{equation}

\begin{figure*}
  \centering
  \includegraphics[width=\textwidth, angle=0]{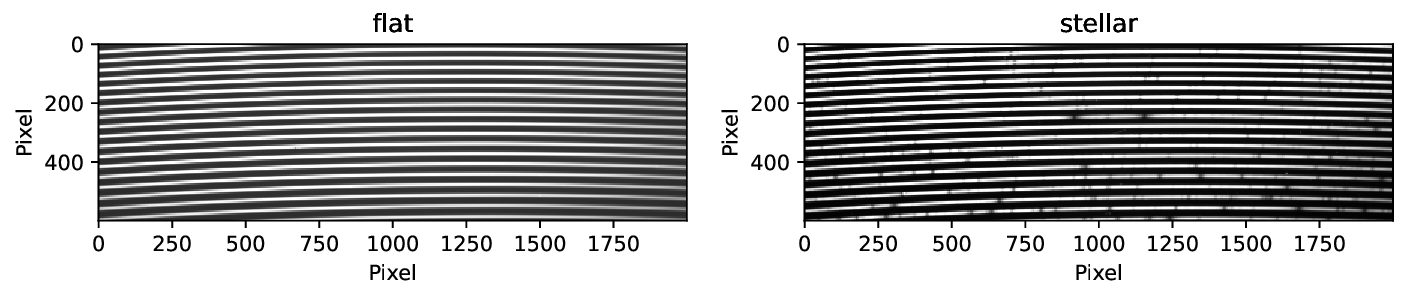}
  \caption{Two-dimensional images of a flat and a star, respectively. The left is flat, while the right is a star (HD189733).}
  \label{f2star}
  \end{figure*}

\begin{figure*}
  \centering
  \includegraphics[width=0.8\textwidth, angle=0]{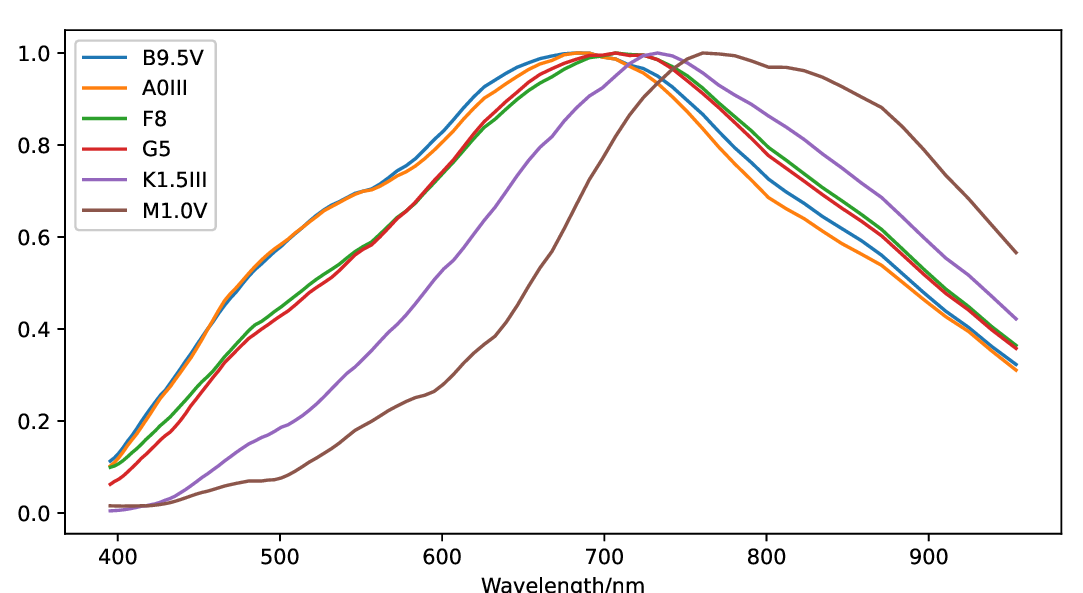}
  \caption{The distribution of ADU (normalized) in different wavelengths observed by stars of different spectral types on CCD (2000th column).}
  \label{bafgkm}
\end{figure*}

 The total count of the PMT is the sum of the exposure fluxes of all the bands it detects. The energy distributions of stars of different spectral types are also different, and with the inconsistent response of HiRES in different bands, as well as the actual atmospheric absorption, it is difficult to obtain more accurate spectral response curves using the standard blackbody radiation formula. Fig. \ref*{bafgkm} shows the actual observed energy distributions for stars of six spectral types, where the observed energies have been normalized. In addition, the actual observed stellar spectra are different from those of the flat field lamps as shown in Fig.~\ref*{f2star}. There are many absorption lines in star's spectrum, and the right image shows that the darker parts of the discontinuity are absorption lines. Based on these spectral characteristics of HiRES, we need to determine the respective conversion factor G matrices for the stars of different spectral types.

\begin{figure*}
  \centering
  \includegraphics[width=0.6\textwidth, angle=0]{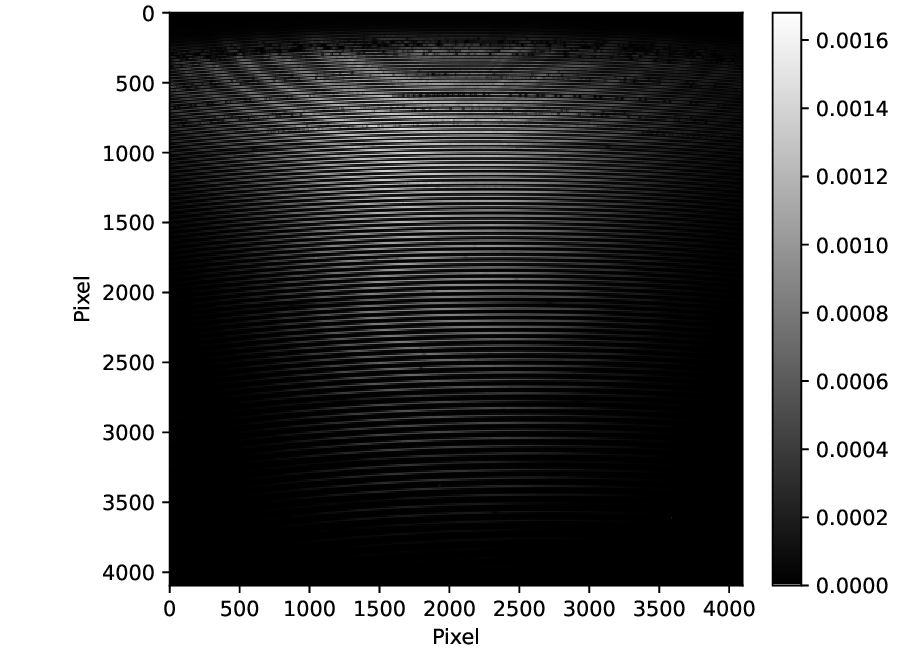}
  \caption{The conversion coefficient matrix for spectral type A.}
  \label{G}
\end{figure*}
With the equation \ref*{eq1}, we determined and obtained six types (Type B, A, F, G, K, and M) of conversion factors, using historical observations. These observations include: CCD spectral image, PMT total counts, CCD dark, and PMT dark counts. Fig. \ref*{G} shows the matrix of conversion coefficient for a star of spectral type A. With the above method, we can estimate the exposure flux of the target spectrum using the real-time count of the PMT and obtain a simulated CCD spectral image.

\section{Validation}
\label{Validation}

We have validated the above method by observations from July 16, 2022 to July 20, 2022. The information of the observed targets is given in Table \ref*{Tab1}, and the CCD dark and PMT dark counts were obtained from the observations on the same day. The conversion factor (G) corresponding to the spectral type of the target to be validated was chosen, and the flux ($F_{ccd}$) of the CCD was estimated by equation \ref*{eq1} using the counts of PMT obtained from the observation. The results were evaluated using a coefficient of determination between the estimated value of the CCD exposure flux and the actual observed value. The coefficients of determination were obtained using equations 2, 3, and 4.

\begin{equation}\label{eq2}
  R^{2}=1 - \frac{SS_{estimated}}{SS_{Total}} 
\end{equation}
\begin{equation}\label{eq3}
  SS_{Total}= \sum_{i=1}^{n}(y_{i}-y_{mean})^2
\end{equation}
\begin{equation}\label{eq4}
  SS_{estimated}= \sum_{i=1}^{n}(y_{i}-y_{estimated})^2
\end{equation}

where $R^{2}$ is the coefficient of determination, $y_{i}$ is the observed flux of the CCD, $y_{estimated}$ is the estimated flux of the CCD, and $y_{mean}$ is the average of the observed fluxes. If the estimate and the observation are in perfect agreement, the value of $R^{2}$ will be 1. The third column of the table \ref*{Tab1} shows that all $R^{2}$ values are greater than 0.92, indicating a high degree of agreement.

\begin{table*}[h]
  \begin{center}
  \caption[]{Information on observed targets for validation.}
  \label{Tab1}
   \begin{tabular}{clclcl}
    \hline\noalign{\smallskip}
   spectral type of stars for validation &  magnitude & $R^2$ \\
    \hline\noalign{\smallskip}
    B9.5V             &  5.665mag. & 0.964\\
    A1V               &  3.7mag.   & 0.970\\
    F8                &  7.58 mag. & 0.964\\
    G1IV              &  7.457mag. & 0.967\\
    K1.5III           &  -0.05mag. & 0.923\\
    M1.0V             &  10.03mag. & 0.990\\
    \noalign{\smallskip}\hline
  \end{tabular}
  \end{center}
  \end{table*}

  \begin{figure*}[h]
    \centering
    \includegraphics[width=0.80\textwidth, angle=0]{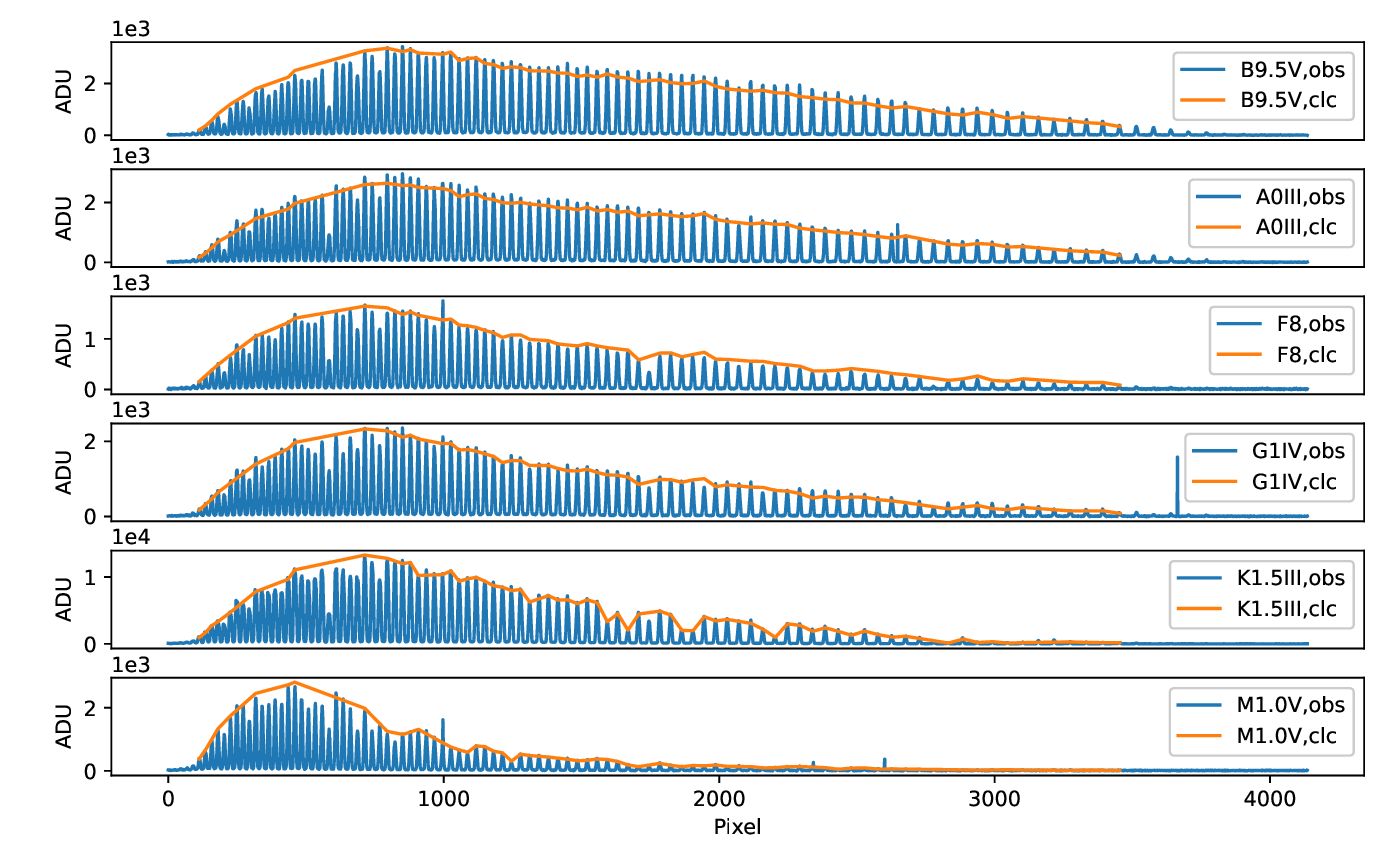}
    \caption{Comparison of estimated and observed values of flux for stars with different spectral subtypes. The blue curves are the observations, and the orange curves are estimation.}
    \label{validation}
    \end{figure*}

Fig. \ref*{validation} shows a plot between the estimated and observed values of the exposure fluxes on pixels of the CCD, where the blue curve represents the observed values and the orange curve represents the estimated values. The absorption line and the region between the two orders in the estimated values are ignored for clarity of comparison. By comparing the results of the above figure and table, it can be concluded that our method can estimate the exposure flux of the CCD using the real-time counts of the PMT in the observation, and the estimation results are reliable.

\section{Software}
\label{soft}

We developed a software that uses the EM to estimate the exposure time based on Python and the Qt Designer IDE of the Windows 10 system. 
\begin{figure*}[h]
  \centering
  \includegraphics[width=0.60\textwidth, angle=0]{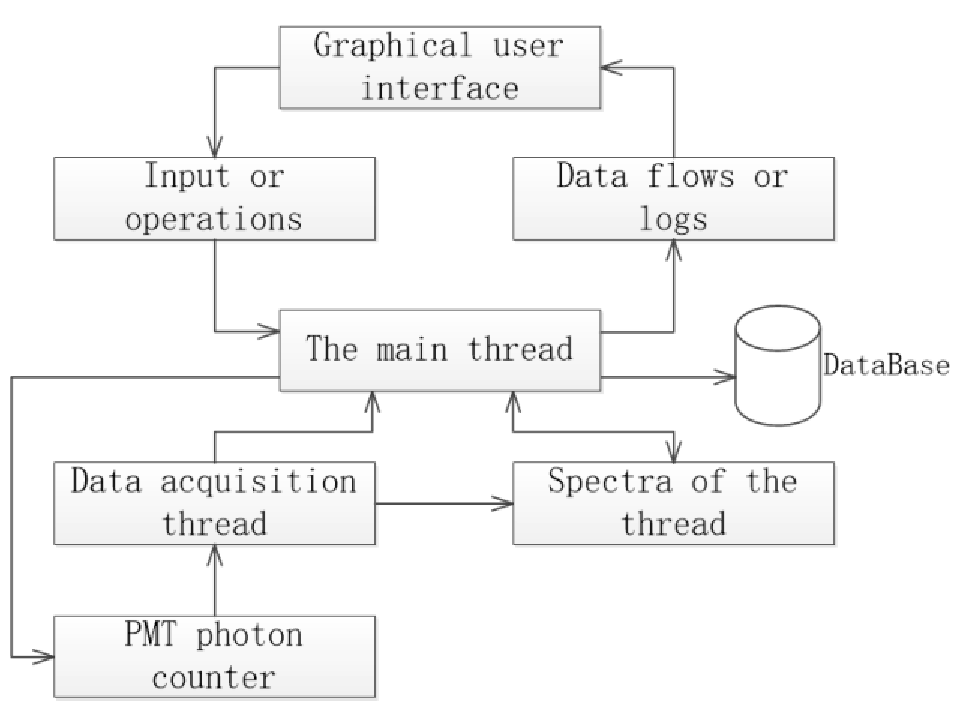}
  \caption{Flow chart of observation software. }
  \label{flowchart}
  \end{figure*}

  \begin{figure*}[h]
    \centering
    \includegraphics[width=\textwidth, angle=0]{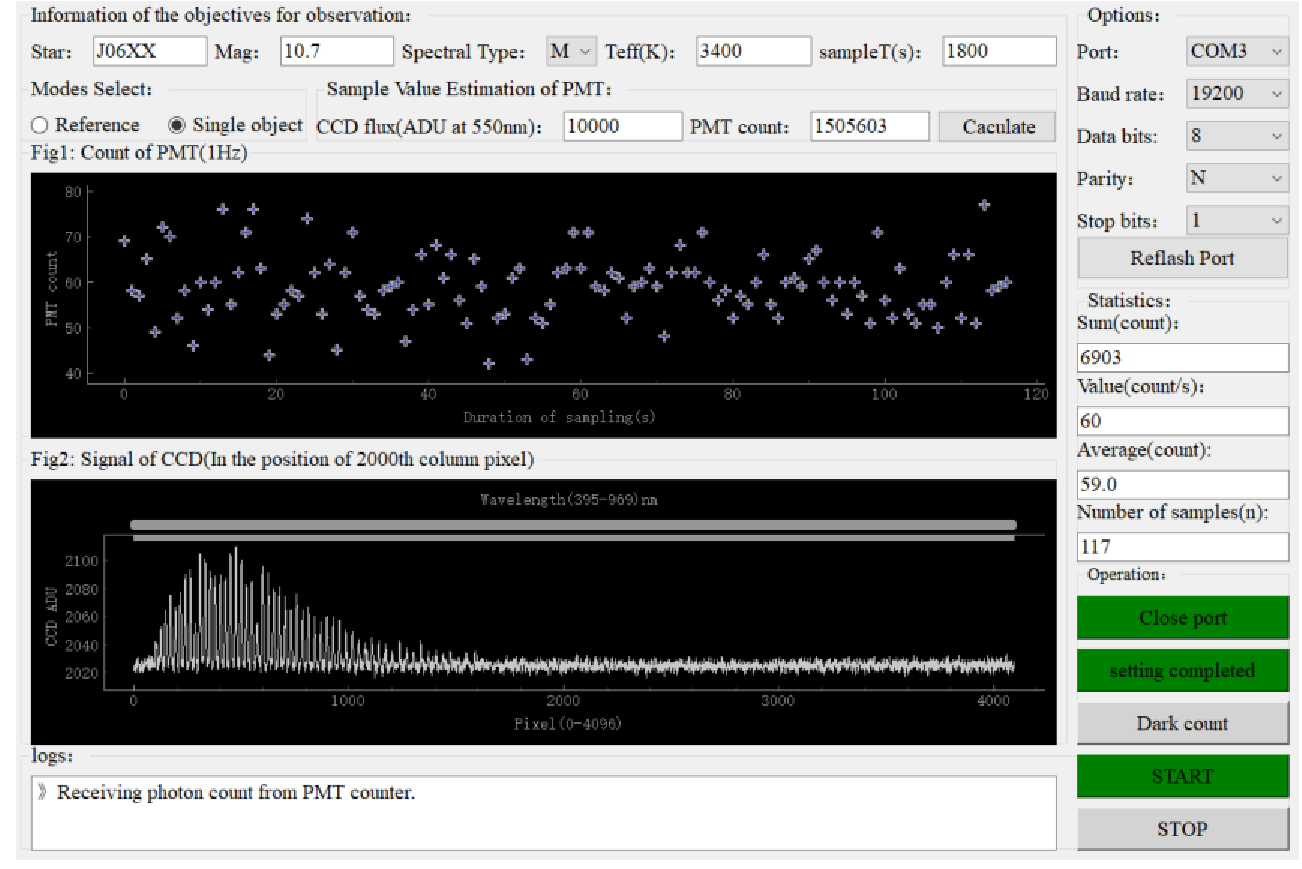}
    \caption{Observation software operation interface. }
    \label{hmi}
    \end{figure*}

Fig. \ref{flowchart} shows the software workflow. The working mechanism is mainly done through the communication of three threads. These threads are the main thread, the data acquisition thread and the spectra of the thread. The main thread receives the operations and sends the command to the PMT counter through the serial port. The PMT counter does as commanded and sends data back to the software. The data acquisition thread is responsible for receiving counts, verifying the data, and subtracting the background counts. The main thread receives the photon counts, computes statistics (including sample value, mean value, total value, and number of samples), and dynamically updates the statistical results. The spectra of the thread is responsible for estimating the current exposure flux at a given frequency (0.2Hz) and updates in real time.

Fig. \ref{hmi} shows the user interface of the software. Observers can select or fill in the star's information in the operation interface according to the scientific requirements before observation. The PMT counts and the estimated exposure flux can be displayed in real-time during observation, providing a visual and reliable reference for the observation.

\section{CONCLUSIONS AND DISCUSSION}
We have presented a method for estimating the exposure flux of a CCD in real-time using the counts of the PMT of an exposure meter,  and show the validation of the method using observational data.  The validation results show that the estimates obtained using the method are very much in line with the actual observations.

A linear and proportional relationship between the total counts of the PMT and the fluxes of the pixels corresponding to different wavelength points of the CCD spectral image has been determined by flat-field lamp observations, and the equation of the transformation relationship is given. The six spectral types' transformation coefficient templates were determined from historical observations. The validation of the observation data from July 16 to July 20, 2022 shows that the coefficient of determination values between the estimated and observed values of the CCD exposure fluxes are all greater than 0.92. It is shown that the method can be used to estimate the CCD exposure flux using real-time counts of the PMT in the observation, and the results are reliable. In addition, the working mechanism and user interface of the method's application software are shown. The software displays in real-time the PMT counts and the estimated value of the exposure flux at the CCD at that time, providing a visual reference for optimizing  exposure time control.

In Section 2, we determine a template of model transformation coefficients for different spectral types using historical observations, and the key lies in the choice of data. First, the data were chosen to cover as many subtypes of the spectral type as possible. Due to precious observing time, we could not observe targets of all subtypes. Second, higher quality observations were selected, such as those with high signal-to-noise ratios under stable observing conditions. The distribution of PMT observations can be used to determine whether the observing conditions were stable at that time and whether they were affected by cloudiness. Considering the above factors, the actual amount of selected observations is small. Third, there are different magnitudes for stars of the same spectral type. We found that the response distribution of spectral imaging on the CCD varies between targets with large magnitude differences, especially in the blue-end wavelength region, due to the atmospheric window and spectrometer efficiency. Therefore, we must also perform separate model transformation coefficient determinations for particularly bright standard stars of the same spectral type, or for particularly faint special targets.

In addition, the effects caused by the sky-light background and the atmospheric absorption line are not considered in this paper. According to the long-term monitoring statistics of Lijiang Observatory, the sky-light background is between 16.5-21 mag./arcsec$^2$ \citep{XYX}. This may introduce some uncertainties in the estimation of faint stars observed with long exposures. This factor must be taken into account if high accuracy is required. For example, we can make an observation of the sky-light background in the region of the target as a background correction before the observation. To a certain extent, it can reduce the influence of the sky-light background, but it also loses additional observation time and affects the observation efficiency. In addition, the intensity of the atmospheric absorption line of Lijiang Observatory varies with the seasons \citep{lkx}, and the flux at the wavelength it covers will also follow the influence, and the observer can avoid referring to the pixel area covered by this wavelength.

Future plans include increasing the sample size of observations to optimize model transformation coefficients to improve estimation accuracy, and improving software functionality based on observation requirements. This research provides valuable insights for the application of the method on other spectrometers of the same type. It should be noted that the current implementation of the method is limited to bands with wavelength coverage less than 10,000\,\AA\ in HiRES for the 2.4m telescope. However, with the selection of an appropriate detector, this method can be extended to other wavelength spectrometers.
\begin{acknowledgements}
This work was funded by the National Natural Science Foundation of China (NSFC, Nos. 11803088, 12003068, 12063002), Civil Aerospace pre research project D020302 and the science research grants from the China Manned Space Project with NO.CMS-CSST-2021-B10, Yunnan Science Foundation of China (202001AU070077). Sino-German Scientist Mobility Programme M-0086.
\end{acknowledgements}






%
\label{lastpage}

\bibliographystyle{raa} 
\bibliography{reference}

\end{document}